\pdfoutput=1

\documentclass{cimentorr}
\usepackage{graphicx}  
\title{Galaxies as a cosmological test}
\author{P. J. E. Peebles}
\instlist{\inst{}Joseph Henry Laboratories, Princeton University, Princeton, NJ USA}
\PACSes{\PACSit{98.62.Ai}{Origin and evolution of galaxies}
\PACSit{98.80.-k}{Cosmology}}
\begin{document}

\maketitle

\begin{abstract}
The relativistic $\Lambda$CDM cosmological model has passed a demanding network of tests that convincingly demonstrate it is a useful approximation to what happened back to high redshift. But there are anomalies in its application to structure formation on the scales of galaxies that show we have much to learn about what this theory actually predicts and possibly something also of value to learn about the fundamental theoretical basis for observational cosmology. 
\medskip

This is slightly revised and enlarged from a contribution to  {\bf A Century of Cosmology}, 
Venice, August 2007.
\end{abstract}

\section{Introduction}
The completion of a tight network of tests of the $\Lambda$CDM cosmology is a remarkable and important advance. But does it signify completion of the fundamental physics that will be needed in the analysis of this and future generations of observational cosmology? Or might we only have arrived at the simplest approximation we can get away with at the present level of the evidence? The history of physical science offers precedents for the latter: advances tend to be followed in due course by deeper advances. 

The general acceptance of $\Lambda$CDM as the working standard is good strategy: if this cosmology is not an adequate approximation to reality it will become manifest in anomalies. One may also to seek alternative physics \cite{ref:DGP} with deeper roots that may fit the present cosmological tests as well as or eventually better than $\Lambda$CDM. This too has a proud tradition: Einstein's argument that a reasonable universe is homogeneous took no note of what astronomers may have told him about how the nearby stars are distributed. Yet another philosophy takes a proactive empirical line: seek anomalies, and search for them on the possibly more fertile ground of the galaxies, below the scales of the present successful cosmological tests. I discuss here two issues in galaxy phenomenology that seem to me to be particularly interesting challenges to standard ideas. 

\section{Void dwarfs}

\begin{figure}[t]
\begin{center}
\includegraphics[width=5.in]{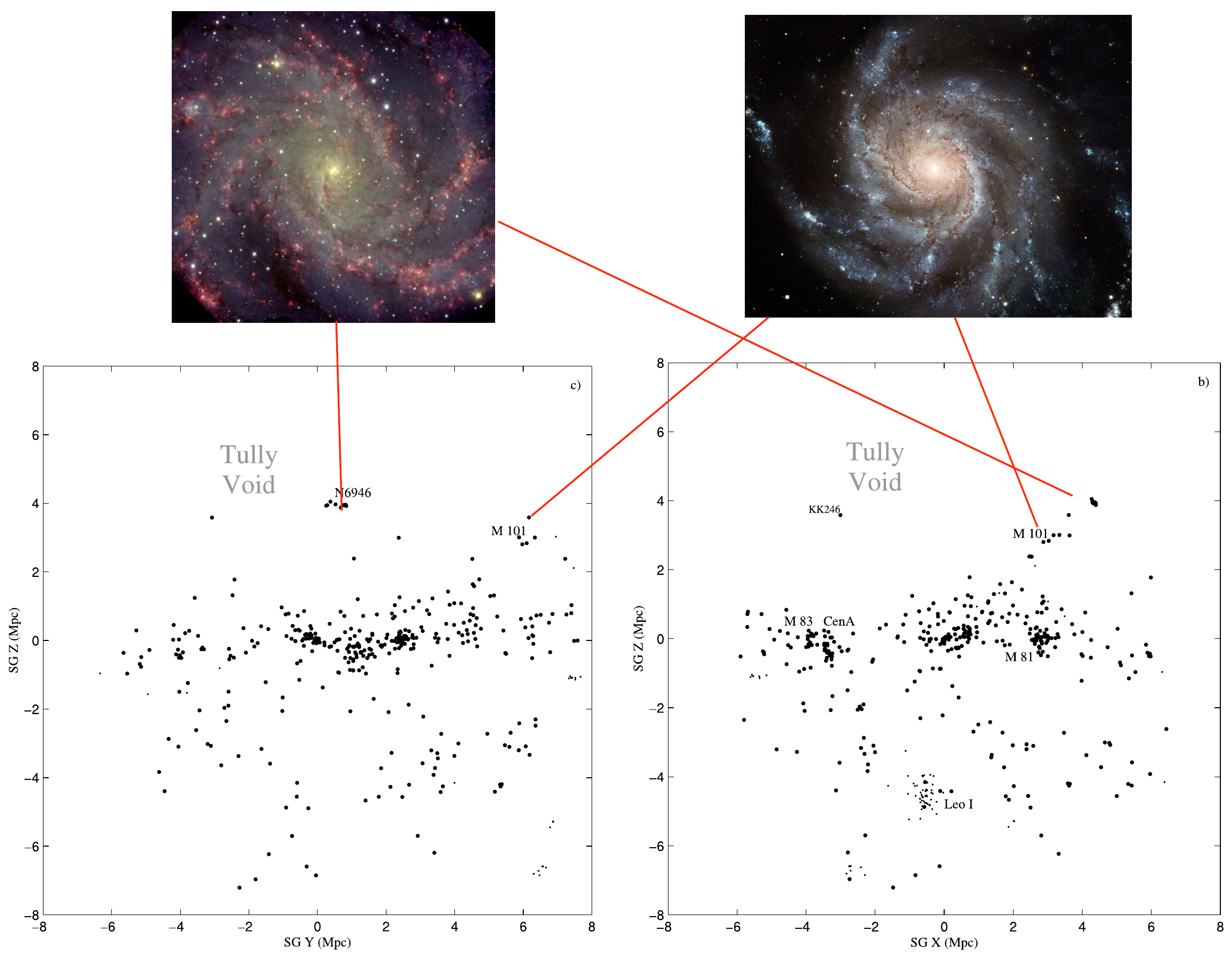} \caption{The Karachentsev {\it et al.} {\it Catalog of Neighboring Galaxies} \cite{ref:Karachentsev}. Two of the largest in this sample, M\,101  and NGC\,6946,  are on the edge of the Local Void, a curious situation.\label{fig1}} 
\end{center}
\end{figure}

The {\it Catalog of Neighboring Galaxies} by Karachentsev, Karachentseva, Huchtmeier and Makarov \cite{ref:Karachentsev} gives us a fascinating example of how the galaxies are   distributed. The map of these galaxies in Figure \ref{fig1} is taken from their Figure~5. The two projections shown here are normal to the plane of the de Vaucouleurs Local Supercluster, and one sees the striking continuation of this planar concentration to our immediate neighborhood. I have added images of two of the largest galaxies in this sample, M\,101 (HST Legacy Image) and NGC\,6946 (Gemini Observatory/Travis Rector, U of Alaska), along with lines to indicate their positions. These two large spirals are in islands edging into Tully's Local Void. We are at the center of the map, and still close to the void. 

How does this situation compare to what might be expected in the $\Lambda$CDM cosmology? Pure cold dark matter numerical simulations show that the matter gathers into mass concentrations --- halos --- that seem to be  suitable homes for galaxies. These halos appear in concentrations outside of which the low density regions look like good candidates for the voids in the observed galaxy distribution. Trails of low mass halos run into the candidate voids regions, while the more massive halos avoid the edges of voids: they prefer regions of high local density. This follows from the assumption of an initially scale-invariant Gaussian random mass distribution. It agrees with the observation that the most massive galaxies, with luminosities $L\simeq 10L_\ast$, prefer the densest regions. 

In the nearby universe of galaxies, where the greatest luminosities are closer to $L_\ast$, the $\Lambda$CDM simulations do not offer such a reasonable-looking account of what is observed. Figure \ref{fig1} shows that two of the largest galaxies in this sample are edging into the Local Void. Apart from the islands of dwarfs around these two spirals there are no known galaxies of any size in the void. (Karachentsev{\it et al.} \cite{ref:Karachentsev} note that KK\,246, which  is labeled in the upper left side of the right-hand projection in Fig.~\ref{fig1}, ``is situated just at the edge of the Tully void, but not inside it.'') These observations do not seem to be a natural outcome of galaxy formation in the standard cosmology.

 \begin{figure}
\begin{center}
\includegraphics[width=3.75in]{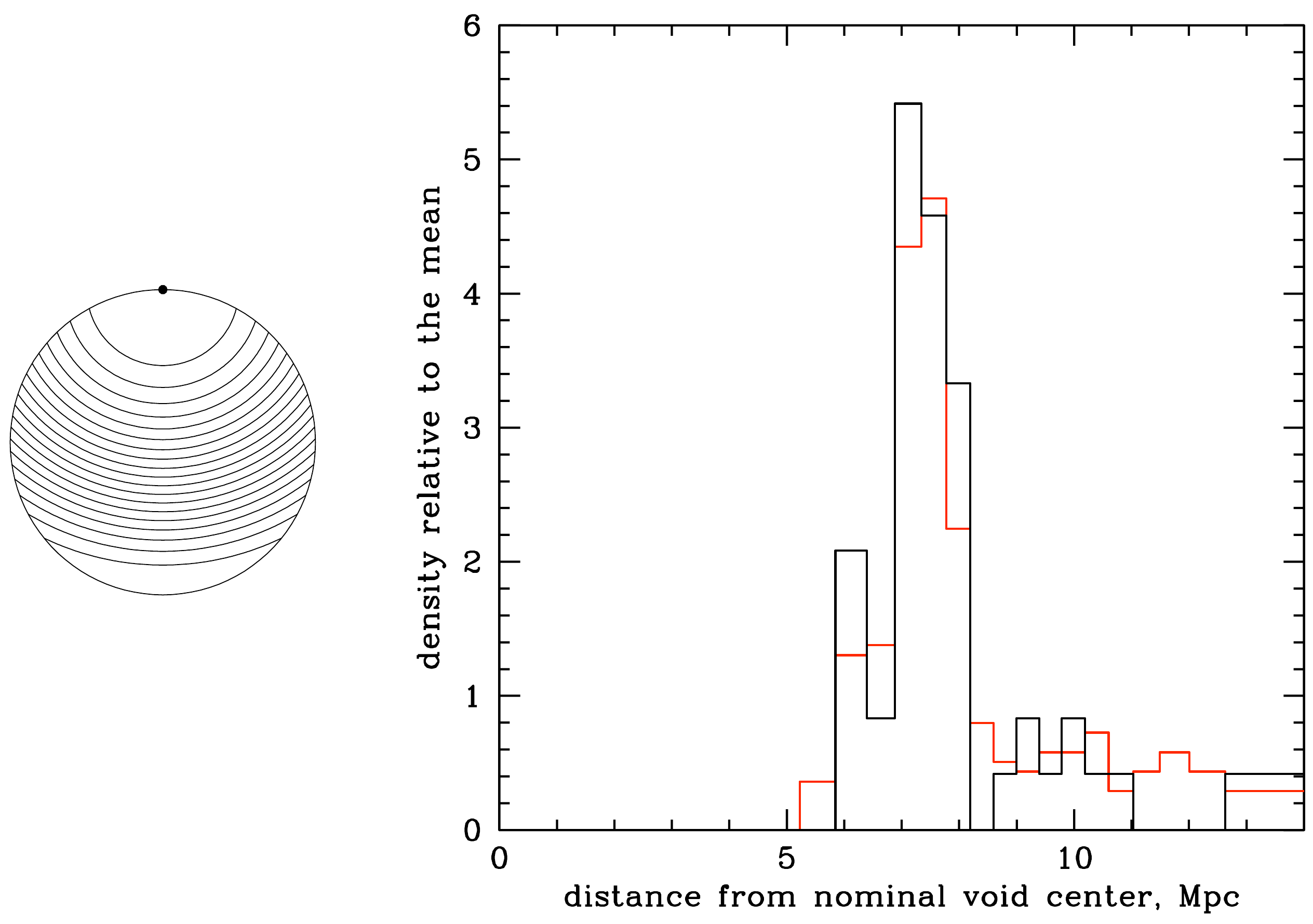}
\caption[]{Number density of galaxies as a function of distance from a nominal center of the Local Void in the 20 spherical slices of equal volume illustrated at the left. The black line shows the run of number density of more luminous galaxies, $M_B<-17$, relative to the mean in the sample, and the red line shows the run of density relative to the mean for dwarfs at $M_B > -17$. \label{fig2}}
\end{center}
\end{figure}

Figure \ref{fig2} shows the situation in a little more detail, in a crude illustration of the relative distributions of large and small galaxies as a function of distance from the Local Void. It is based on the sample of  Karachentsev {\it et al.} \cite{ref:Karachentsev}  galaxies at distances $<7$~Mpc. The space within the 7~Mpc sphere centered on us is sliced into 20 equal volumes by a nested set of spheres centered on the edge of the 7~Mpc sphere, at right ascension $\alpha = 265^\circ$ and declination $\delta =15^\circ$. (This is the angular position that maximizes the distances to the nearest galaxies within the 7~Mpc sphere). The equal volume slices are illustrated on the left side of Figure~\ref{fig2} (and recall that this is a figure of revolution). 

The galaxy count $N_i$ in the $i^{\rm th}$ slice is converted to the ratio of the mean number density $n_i$ within this slice to the mean number density $\langle n\rangle$ in the whole 7~Mpc sphere by 
\begin{equation}
n_i/\langle n\rangle = 20N_i/N,
\end{equation}
where  $N$ is the total number of galaxies in the sphere. These number densities, computed separately for the dwarfs at $M_B > -17$ and for the more luminous galaxies at $M_B < -17$, are plotted as the red and black lines respectively in Figure \ref{fig2}. 

I am impressed by the very similar distributions of dwarf and more luminous galaxies. Both avoid the void (which is represented in too few empty slices because the slices are placed in a clumsy way). The peaks roughly mark the concentration of galaxies in the plane of the Local Supercluster. Similar fractions of the giants and dwarfs are in this concentration. Below the plane the number densities of both are about half the mean within the sphere. This does not seem to be a natural outcome of the standard cosmology.

A third way to look at the situation is based on halo mass functions. Warren,  Abazajian,  Holz and  Teodoro \cite{ref:warren} find that in pure CDM numerical simulations the  low mass end of the halo mass function is well approximated by the power law
\begin{equation}
n(>m) = 2.1\, h^3\left(10^9h^{-1}\hbox{ Mpc}\over m\right)^{0.9}\hbox{ Mpc}^{-3}.
\label{eq:Warren}
\end{equation}
Hoeft, Yepes, Gottl{\"o}ber and  Springel \cite{ref:Hoeft} find that in voids the low mass end is well approximated as 
\begin{equation}
n(>m) = 0.2\, h^3\left(10^9h^{-1}\hbox{ Mpc}\over m\right)^{1.1}\hbox{ Mpc}^{-3}.
\label{eq:Hoeft}
\end{equation}
The former is computed to halo mass $m=3\times 10^{10}m_\odot$; the close approximation to a power law invites the extrapolation to extreme dwarfs at $m\sim 10^{9}m_\odot$. Hoeft {\it et al.} caution that the number density of extreme dwarf halos is larger near the edge of a void than near the center. Since the Local Void in the region sampled by the {\it Catalog of Neighboring Galaxies} is small, equation~(\ref{eq:Hoeft}) is not likely to underestimate the predicted numbers of extreme dwarfs in this region. 

Equations~(\ref{eq:Warren}) and~(\ref{eq:Hoeft}) say that the number density of extreme dwarfs in voids is about 10\% of the global number density, a number that is often mentioned. The right-hand map in Figure \ref{fig1} suggests the Local Void occupies about one third of the region within 7~Mpc. Thus if efficiency of conversion of CDM halos to galaxies observable in optical or H{\small I} emission were independent of environment then we would expect that about one in thirty of the members of the {\it Catalog of Neighboring Galaxies} is in the Local Void. There are 29 galaxies brighter than $M_B=-18$ at distances $1<D<7$\ Mpc (which excludes the Local Group). This calculation predicts that about one of them is in the Local Void, which is no problem. There are 250 galaxies at $-18<M_B<-10$ and $1<D<7$~Mpc. The calculation predicts about ten are in the Local Void. So where are they? 

This is based on pure CDM simulations, not the Halo Occupation paradigm. The extrapolation of the global mass function in equation~(\ref{eq:Warren}) does not seem dangerous but a check of that and of the halo mass function in a low density region with the small size of the Local Void would be welcome.

Karachentsev {\it et al.} \cite{ref:Karachentsev} caution that their sample is not likely to be complete at $M_B=-10$. But completeness is not a manifest problem because we are comparing counts of dwarfs in regions of high and low density at about the same distance and hence perhaps similar discovery efficiency. 

In this statistic there is no problem with the absence of galaxies as luminous as the LMC well inside the void. (The similar distributions of giants and dwarfs as functions of distance from the local void in Fig.~\ref{fig2} is a separate issue.) The problem is the absence of void objects at luminosities $M_B\sim -13$ characteristic of NCG\,3741 and DDO\,154, both at about 3\,Mpc distance. Both have elegant H{\small I} disks \cite{ref:Begum}, \cite{DDO154}. It would be fascinating to see examples of such objects in voids, at distance greater than a few megaparsecs from the nearest normal galaxy, and to learn how common they are. 

The more widely discussed problem with dwarfs in simulations is that their numbers in halos of large galaxies are much larger than the observed numbers of satellites. The usual interpretation that most dwarf halos have not retained observable amounts of stars or H{\small I}. Might void dwarf halos be even less likely to retain observable amounts of baryons than dwarf satellites? I think the most direct objection to this hypothesis is that gas-rich extreme dwarfs are seldom seen closer to an $L_\ast$ galaxy than a few hundred kiloparsecs \cite{ref:Geha}. That suggests an $L_\ast$ galaxy suppresses retention of gas in dwarfs in its dark halo. And that suggests dwarf halos are more likely to be visible in low density regions. If so it  would exacerbate the problem with the absence of extreme dwarfs in the Local Void. 

One may ask why I am so concerned about only ten missing dwarfs in the Local Void: maybe it's a statistical fluke; maybe the local situation is anomalous. It is fortunate that the ALFALFA \cite{ref:alfalfa} survey at the Arecibo radio telescope is capable of showing us the distribution of H{\small I}-rich dwarfs in and around other nearby voids. ALFALFA is addressing what may prove to be a fundamental issue for the physical basis of cosmology: how are gas-rich voids distributed relative to $L_\ast$ galaxies in and around low density regions? 

\begin{figure}[t]
\begin{center}
\includegraphics[width=5.25in]{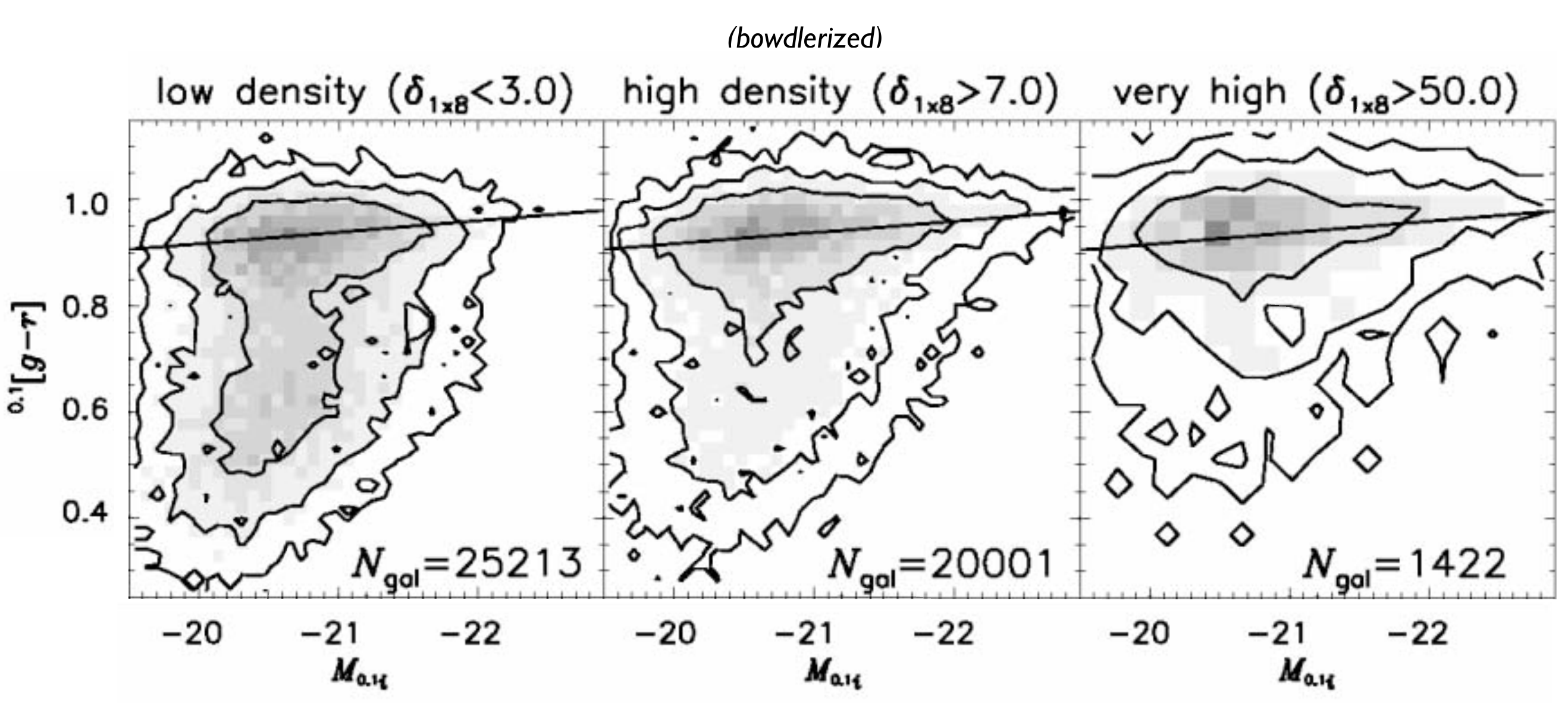} 
\caption{Color-magnitude diagrams of SDSS galaxies as a function of ambient density \cite{ref:Hoggetal}.\label{fig3}} 
\end{center}
\end{figure}

\section{Giant galaxies at low redshifts}

The most massive and luminous galaxies favor regions of high ambient density, as predicted by the $\Lambda$CDM cosmology. But there is a problem.  Observations suggest that giant galaxies have evolved in a good approximation to island universes since redshifts greater than unity. That does not naturally agree with the indication from numerical simulations that at redshifts less than unity a giant galaxy typically has exchanged considerable amounts of matter with its surroundings out to distances of megaparsecs.  

The example of the island universe behavior in Figure~\ref{fig3}  is excerpted from Figure~1 in Hogg {\it et al.} \cite{ref:Hoggetal}. The three panels show distributions in galaxy color (bluer toward the bottom, redder toward the top) and absolute magnitude (the luminosity increases to the right). The $\sim 50,000$ galaxies in this statistical sample from the Sloan Digital Sky Survey are separated into three subsets by ambient density. (The density estimate necessarily is noisy --- Park {\it et al.} \cite{ref:Park} present another estimator ---  but comparison with what they find shows that that does not affect the statistical relations illustrated in this figure.) 

Figure~\ref{fig3} is quite informative. The galaxies in the ``blue plume'' extending downward near the left side of each panel contain more interstellar gas that is producing more young luminous blue stars than the red galaxies that are placed toward the top of the figure. The fraction of galaxies in this blue plume increases with decreasing density: gas-rich galaxies prefer low ambient density. The solid lines toward the top of each panel mark the ``red sequence'' of early-type galaxies, largely ellipticals and S0s, that contain relatively few young blue stars. The red sequence extends to the right of the blue plume: the most luminous galaxies are on the red sequence. The red sequence extends furthest to the right at the highest ambient density: the most luminous galaxies prefer the densest environments.

\begin{figure}[t]
\begin{center}
\includegraphics[width=3in]{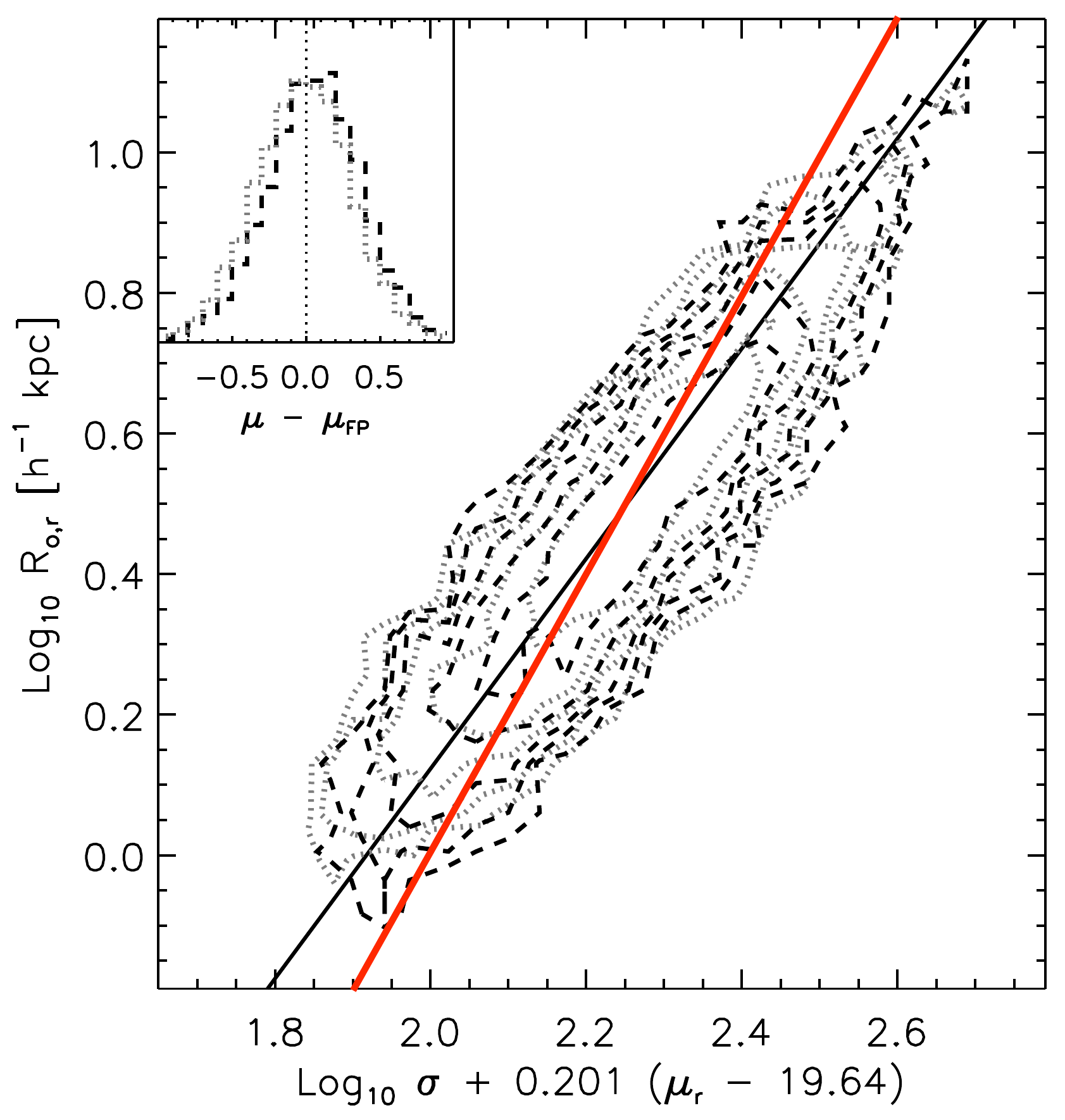} \caption{Bernardi {\it et al.} \cite{ref:Bernardi} distributions of early-type galaxy radius $R$, stellar velocity dispersion $\sigma$ and surface brightness $\mu$. Dashed contours represent the higher ambient density subset, dotted contours the lower ambient density. The added line  in red assumes constant  mass-to-light ratio. \label{fig4}} 
\end{center}
\end{figure}

The final point, and the one of direct interest here, is the color-magnitude relation represented by the solid line. The line is the same in each panel. It is difficult to see any difference among the distributions of red galaxies around this line in the three subsets. There are tighter demonstrations of this color-magnitude relation in nearby galaxies, but they tend to measure colors near the centers of galaxies, which need not be representative. Figure~\ref{fig3} shows colors of most of the starlight. And it demonstrates that the luminosity of a red galaxy is a good predictor of its color wherever the galaxy finds itself. That is,  the red sequence galaxies have the appearance of island universes: their properties are more strongly determined by their stellar mass than by their environment.

We have another illustration from the fundamental plane correlation among galaxy radius, luminosity and stellar velocity dispersion. In Figure~\ref{fig4}, from Bernardi {\it et al.} \cite{ref:Bernardi}, I have added a red line to indicate the relation predicted from the virial theorem, $GM\sim R\sigma^2$, and constant mass-to-light ratio: $\log\sigma +0.2\mu = 0.5\log R +\hbox{ constant}$. Dashed contours represent galaxies in higher density environments, dotted lower ambient density. In both, $M/L\propto R^{0.3}$. Here radius, not environment,  predicts the galaxy mass-to light ratio. Again, the appearance is of island universes.

Environment certainly matters: we have good evidence that that determines the distribution of halo masses. And at given halo mass early-type galaxies are not entirely indifferent to their environment: the excellent statistics in their sample allow Bernardi {\it et al.} to demonstrate the difference between high and low density subsets indicated in the inset in Figure~\ref{fig4}. But this is is a wonderfully small effect. Also wonderfully small is the sensitivity to environment of  the ``frosting'' of younger stars in large galaxies, which is observed more commonly in lower density regions \cite{ref:Faber}. I rank this near indifference to environment high on the list of criteria for a satisfactory theory of galaxy evolution. And on this score I see a real problem with $\Lambda$CDM, as follows. 

\begin{figure}[t]
\begin{center}
\includegraphics[width=5.25in]{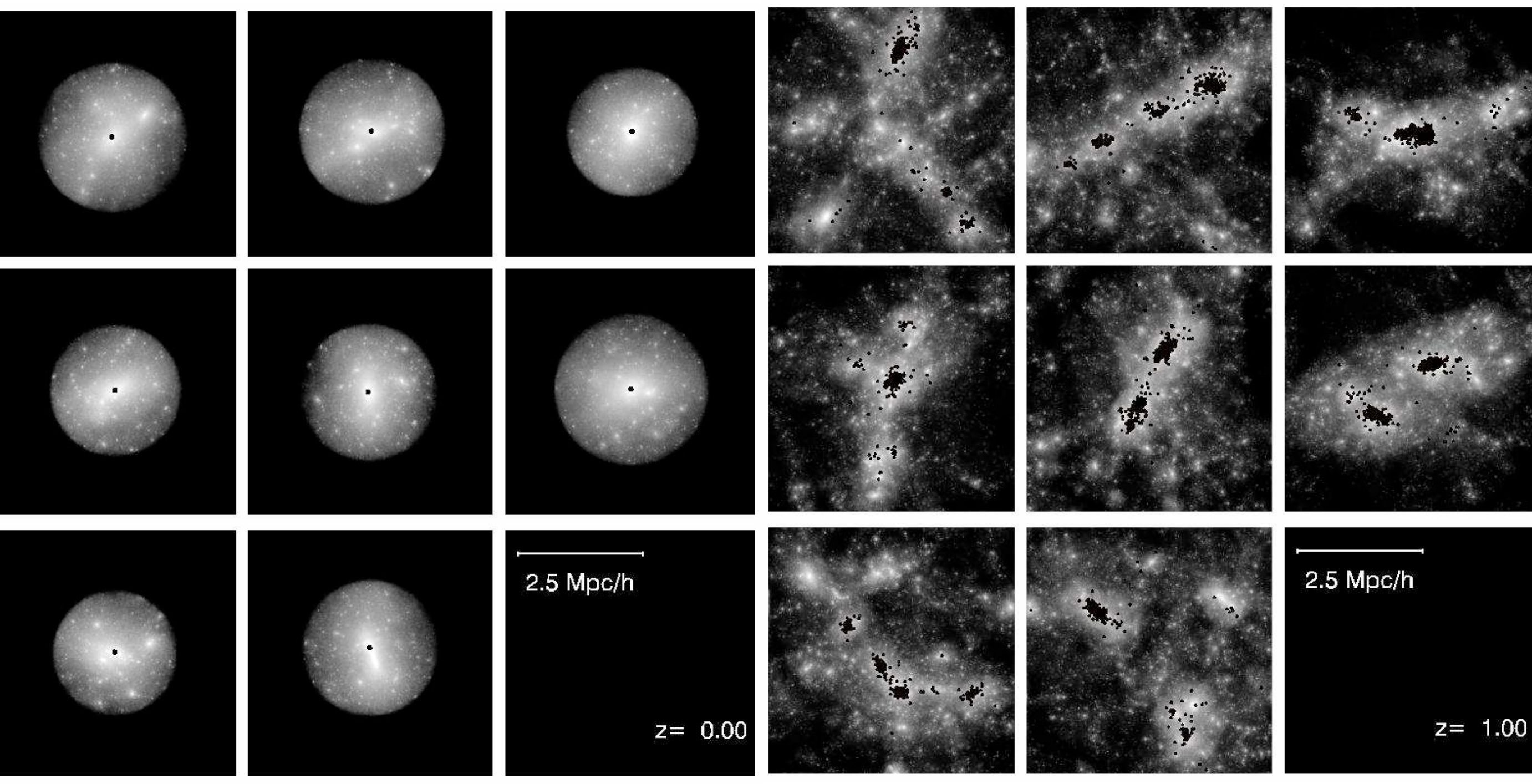} \caption{Evolution from $z=1$ to the present of the mass distribution in and around eight examples of giant galaxy halos in a pure CDM simulation \cite{ref:Gaoet al}. Particles that are within $15$~kpc of the center of the halo at the present epoch are plotted in black in both panels. Particles that are outside this radius and within $r_{200}$ at the present epoch are plotted in white. The rest of the matter is not shown.\label{fig5}} 
\end{center}
\end{figure}

The simulation of cosmic evolution shown in Figure~\ref{fig5} has $\Lambda$CDM  parameters except that there are no baryons: the matter is pure CDM \cite{ref:Gaoet al}. The only particles plotted are the ones within the nominally virialized --- stable --- parts of the most massive halos at the present epoch. These halos are likely hosts of the most massive galaxies. The black circle at the center of each massive halo at $z=0$, in the left-hand panel, has radius 15~kpc (at the standard distance scale). This is a reasonable approximation to the characteristic stellar radius of a giant galaxy. The positions of the particles that are inside this radius at the present epoch are plotted in black in both panels. The black markers are visible in the right-hand panel, at $z=1$, because they are surrounded by the particles plotted in white that  at $z=0$ are inside a conventional approximation to the present virial radius and outside 15~kpc. The length scales are physical; each box width is 7~Mpc. 

A remarkable --- and important --- regularity is that in each of the regions shown in the figure the largest mass within a 15~kpc sphere is close to constant all the way back to $z\sim 8$. Matter flows into this sphere, and matter flows out. What is more, the identity of the most massive sphere occasionally changes.  But the largest mass contained remains close to constant. This is important because it means there is no manifest problem with the existence of massive galaxies at high redshifts. 

The problematic effect the figure shows us is that in the $\Lambda$CDM cosmology large galaxies ought to show effects of the exchange of matter with surroundings that at redshifts $z<1$ can be quite variable on the scale of the figure. How did the most massive galaxies know to gather preferentially the reddest of the stars in the galaxies that at $z=1$ were scattered over some 5~Mpc, while somewhat less massive galaxies gathered somewhat less red stars? Even more puzzling,  how could this process manage to preserve the indifference to environment one sees in the color-magnitude relation in Figure~\ref{fig3}? How could this gathering of matter from diverse environments preserve the correlation of mass-to-light ratio with radius in Figure~\ref{fig4} that also manifests so little interest in environment? 

How do baryons affect the picture? Let us recall that a star once formed moves  like a dark matter particle. Also, in dark matter merging simulations a particle originally in a dense region tends to end up in a dense region. Stars at high redshift presumably formed in dense regions, and we may expect therefore that old stars would end up toward the dense central parts of a giant galaxy, as observed. That is, heavy merging at $z<1$ does not manifestly violate the radial gradient of the ratio of stellar to dark matter mass densities in galaxies. It would be interesting, and maybe feasible, to check whether the late merging unduly dilutes the stellar mass density in the central parts of a giant galaxy. 

A related issue is the interpretation of the total mass density run in the luminous parts of a large galaxy. The Wu-Tremaine \cite{ref:M87} power law fit for M87 is $\rho\propto r^{-1.8\pm 0.2}$, steeper than the NFW inner profile $\rho\propto r^{-1}$. The difference would not be surprising if giant galaxies were island universes: in the standard model dissipation before and during star formation increases the central baryon density, increasing the gradient of the total mass density. But simulations suggest that merging drives relaxation to a form close to NFW, as in an attractor. Is the amount of merging at $z<1$ in the $\Lambda$CDM model for giant galaxies large enough to have driven relaxation to NFW at $z=0$? That might be checked. In a simulation one might identify ``galaxies'' at $z=2$, say, reshape them to $\rho\propto r^{-2}$  inner parts, and label most of the inner particles ``stars.'' Then compute at $z=0$ the typical mass density and the ratio of star to dark matter densities as functions of radius in ``galaxies.''  This is an easy thought; I don't know how hard it would be in practice. 

\section{Concluding remarks}

I have been complaining about the absence of dwarfs in voids for at least 20 years \cite{ref:Peeb89}. (It  gets hard to remember.) I have not noticed that that has raised much interest in the community. Does that mean I have been  fooling myself? It's happened, but it is not what the evidence on void dwarfs continues to suggest to me. 

I have also long complained about the standard model for giant galaxies, though that has changed. I used to think the argument for late assembly of the largest galaxies is a problem for the standard cosmology \cite{ref:Peebles}. I now realize that the predicted flow of matter into the effective radii of giant galaxies at redshifts $z<1$ is balanced by ejection of matter. My complaint becomes the apparent difficulty of reconciling the promiscuous sharing of matter in giant galaxies shown in Figure~\ref{fig5} with the striking indifference to environment  illustrated in Figures~\ref{fig3} and~\ref{fig4}. So now am I fooling myself? Here too it is not what the evidence seems to be saying to me. 

Galaxy formation and evolution is a complex process. Does this make the phenomenology unsuitable for tests of cosmology \cite{ref:Efstathou}?  Masjedi, Hogg and Blanton \cite{ref:Masjedietal} show a counterexample, a potentially meaningful test. The issues I have discussed here offer other prospects for cosmological tests. If these or other approaches based on galaxy phenomenology show that there is a problem with $\Lambda$CDM then there are ways to fix it without violating the tight network of successful cosmological tests. I tend to like some of the ideas now under discussion  \cite{ref:Farrar}, \cite{ref:Nusser}, but I am most strongly influenced by two general thoughts. We make progress in science by successive approximations. And in physical cosmology, which is data-starved, it makes sense to see whether galaxy phenomenology has something of value to add to the progress.

\acknowledgments

My thinking on these issues has benefitted from ongoing discussions with David Hogg and feedback at colloquia, notably at NYU in New York and the Institute for Astronomy in Hawaii.

\end{document}